\documentclass[5p]{elsarticle}

\makeatletter
\def\ps@pprintTitle{%
 \let\@oddhead\@empty
 \let\@evenhead\@empty
 \def\@oddfoot{\centerline{\thepage}}%
 \let\@evenfoot\@oddfoot}
\makeatother

\usepackage[utf8]{inputenc}
\usepackage[english]{babel}
\usepackage[babel]{csquotes}
\usepackage{stmaryrd}
\usepackage{amssymb}
\usepackage{amsmath}
\usepackage{amsthm}
\usepackage{mathtools}
\usepackage{graphicx}
\providecommand{\l}{\mathit{l}}

\newtheorem{remark}{Remark}

\newcommand{\li}[1]{\ensuremath{{^{(5)}}#1}}

\begin{document}

\begin{frontmatter}
\title{$\Lambda$CDM $\cup$ MOND}

\author{Christian Henke} 
\ead{henke@math.tu-clausthal.de}

\address{University of Technology at Clausthal, Department of Mathematics,\\ Erzstrasse 1, D-38678 Clausthal-Zellerfeld, Germany}

\date{\today}

\begin{abstract}
It has been demonstrated that the difference between the Renormalised Brane World (RBW) model and the Lambda Cold Dark Matter ($\Lambda$CDM) model occurs only at  sufficiently distant times.
In this paper, it is shown that for spherically symmetric situations an analog deviation  between the RBW model and Newton's theory occurs at large distances. 
More precisely, this deviation of the RBW model is nothing other than the explanation of Milgrom's hypothesis and follows from itself. Therefore, the results of this paper explains flat rotation curves of galaxies without dark matter.
\end{abstract}


\end{frontmatter}

\section{Introduction}
Numerous astronomical data and the application of Newton's theory of gravity indicate the existence of mass discrepancies in the universe.
The almost constant rotational speeds of stars at large distances from the galaxy centre can only be explained by Newton's theory of gravity with the presence of additional matter (dark matter). 
Alternatively, these observations could be explained by a breakdown of Newton's theory on galactic scales. 

A phenomenological explanation called Modified Newtonian Dynamics (MOND), which is still very successful today, was given by \cite{Milgrom_Modification_Newtonian_dynamics_1983}.
However, a critical outstanding issue is the development of an acceptable relativistic parent theory for MOND (cf. the discussions in \cite{McGaugh_Tale_two_paradigms2015}).

Attempts to connect MOND to static $3+1$ brane worlds were done in \cite{Milgrom_MOND_theoretical_aspects_2002} and \cite{Milgrom_MOND_from_brane-world_2019}. 
Here, the $3+1$ coordinate for the brane description and the volume of the brane were identified with the non-relativistic gravitational potential and the action of the potential, respectively. The extremisation of this action satisfies the non-linear generalisation of the Laplace equation. With the addition of non-relativistic matter actions, the MOND behavior can be reproduced. The extension to relativistic effests is still an open issue.

Moreover, $4+1$ brane models have been intensively studied for more than two decades. 
The weak gravitational field in the Randall-Sundrum brane world has been well understood (see \cite{Martens.Koyama_Braneworld_Gravity_2010} for an overview) and yields the ordinary Newton potential for a small curvature radius of the Anti de Sitter (AdS) space-time. However, the energy-momentum tensor of the Randall-Sundrum model, which is defined by the extrinsic curvature (compare e.g. \cite{Shiromizu.Maeda.ea_Einstein_3brane_world_2000}), diverges if the brane is moved to the conformal boundary (see \cite{Balasubramanian.Kraus_StressTensor_1999}). 
In contrast, the RBW model is based on a renormalised variational principle and
identifies
the renormalised and therefore finite energy-momentum tensor on the brane with the variation of the matter action (cf. \cite{Henke_Standard_Cosmology_2021}). 
In that work two things have been shown: First, a five dimensional AdS space-time follows from the renormalised variational principle and second, 
the RBW model 
agrees with the exception of the first epoch to an effective $\Lambda$CDM model.  
This agreement can only be achieved if the four dimensional brane of our universe is near the boundary of the AdS space-time. 

The novelty of this work is to demonstrate that the identification of the renormalised energy-momentum tensor with the variation of the matter action provides a natural explanation for dark matter and can be formulated with the MOND model.

In the recent paper \cite{Skordis.Zlosnik.ea_NewRelativisticTheory_2021}, a relativistic gravitational theory was proposed which includes the MOND hypothesis and agrees with the observed cosmic microwave background (CMB). 
While \cite{Skordis.Zlosnik.ea_NewRelativisticTheory_2021} adds new fields resp. new substances to the action, the RBW model doesn't use any additional fields.
Moreover, \cite{Skordis.Zlosnik.ea_NewRelativisticTheory_2021} examines the fluctuations around the cosmological background metric, it does not investigate 
the higher order terms $a^{-s}, s>3$ of the scale factor $a$ in the modified Friedmann equations. In particular, the dynamics in the early universe can cause deviations from the $\Lambda$CDM model.
On the other hand, the study of the linear fluctuations of the RBW model is still open.

The remainder of the paper is organised as follows. In section 2 we investigate the RBW model for static spherically symmetric branes. The next section discusses the non-relativistic limit and the connections to the MOND hypothesis. Finally, section 4 is devoted to concluding remarks.
The general formalism of a $4+1$ decomposition of a five-dimensional bulk space-time can be found in the appendix.



\section{Static spherically symmetric branes}
\label{sec:XXX}
Let $G$ be the metric of the bulk space-time $\mathcal{M}=\mathcal{M}_d \times (y_{\text{br}},\infty)$ with a boundary $\partial \mathcal{M}=\mathcal{M}_d \times \{y_{\text{br}}\}.$ 
Using Gaussian normal coordinates any asymptotically AdS metric can be written in the form 
\begin{equation*}
ds^2=G_{AB}\, dx^A dx^B=\gamma_{\mu\nu}(x,y)\, dx^\mu dx^\nu+dy^2,
\end{equation*}
where the capital Latin letters are used for bulk indices and the Greek alphabet for $d$ space-time indices.
The bulk space-time can be seen as a family of foliated timelike hypersurfaces which are labeled by their coordinate $y.$
In this paper five-dimensional metrics of the form 
\begin{equation}
ds^2=-n(r,y) dt^2+a(r,y) dr^2 +b^2(r,y) d\Omega^2 + dy^2,
\label{eq:RW_metric}
\end{equation}
where the line element of the 2-sphere is given by $d\Omega^2=d\theta^2+\sin^2(\theta) d\phi^2,$ are considered. 

According to \cite{Israel_Singular_hypersurfaces_thin_shells_1966} the following tensor
\begin{equation}
\tilde{T}_{\mu\nu}=- \frac{2}{\kappa_5^2}\left(K_{\mu\nu}-K \gamma_{\mu\nu}\right), \quad K=K_{\mu\nu} \gamma^{\mu\nu},
\label{eq:T_matt}
\end{equation}
can be introduced as a surface energy-momentum tensor, where $\kappa_5^2=8 \pi G_5$ and $K_{\mu\nu}$ denotes the extrinsic curvature tensor (see the definition $(\ref{eq:Kmunu})$ in the Appendix).
Moreover, the intrinsic curvature of $\gamma_{\mu\nu}$ is described by the Riemann tensor
\begin{equation*}
R_{\mu\nu\alpha}^{\phantom{xxx}\beta}=-\left( \partial_\mu \Gamma_{\nu\alpha}^{\phantom{xx}\beta} 
+\Gamma_{\mu\gamma}^{\phantom{xx}\beta} \Gamma_{\nu\alpha}^{\phantom{xx}\gamma}-\mu \leftrightarrow \nu\right),
\end{equation*}
and $R_{\mu\nu}=R_{\mu\alpha\nu}^{\phantom{xxx}\alpha}, R=R_\mu^\mu.$

On the boundary hypersurface $y_{\text{br}},$ the following Einstein equations are satisfied (cf. equations $(\ref{eq:einstein4})$ and $(\ref{eq:en_ten_cft})$ of the Appendix)
\begin{equation}
\begin{split}
R_{\mu\nu}-\Lambda_5 \gamma_{\mu\nu}&=\partial_4 K_{\mu\nu}+\frac{1}{2}\gamma_{\mu\nu} \partial_4 K\\
&+K K_{\mu\nu}-2 K_{\mu}^\alpha K_{\alpha\nu} +\frac{1}{2} \gamma_{\mu\nu}K_{\alpha\beta}K^{\alpha\beta}\\
&\eqqcolon A_{\mu\nu},
\end{split}
\label{eq:brane1}
\end{equation} 
and, if the Bach tensor of the conformal boundary metric is small or vanishing, 
\begin{equation}
\begin{split}
R_{\mu\nu}-\Lambda_5 \gamma_{\mu\nu}&=\kappa_4^2 \left( T_{\mu\nu}-\frac{1}{2}\gamma_{\mu\nu} T \right)\\
&+\frac{\kappa_4^2}{2} \left( \tilde{T}_{\mu\nu}-\frac{1}{2}\gamma_{\mu\nu} \tilde{T} \right)\\
&\eqqcolon B_{\mu\nu},
\end{split}
\label{eq:brane2}
\end{equation} 
where $\Lambda_5=-\frac{6}{l^2},\kappa_4^2=2 \kappa_5^2/l$ and $\partial_4$ is the differentiation w.r.t. $y.$ 
Here, $l$ denotes the length scale of the AdS space-time and $T_{\mu\nu}$ is the renormalised energy-momentum tensor 
 for space-time and matter (cf. the notations of equation $(\ref{eq:not_einstein})$ in the Appendix).
\begin{remark}
If the Bach tensor is not small, then one can argue with the asymptotic expansion near the conformal boundary:   
The closer the brane model is to the conformal boundary, the larger $E_{\mu\nu}-\frac{6}{l^2} \gamma_{\mu\nu}$ is compared to $4 r_{\text{br}}^2\log\left(r_{\text{br}}^2\right) h_{(4)\mu\nu}$ (cf. (\ref{eq:not_einstein})). Moreover, the closer the brane model is to the conformal boundary, the more similar the brane model and the standard model of cosmology are (cf. \cite{Henke_Standard_Cosmology_2021}). Therefore, one can neglect $h_{(4)\mu\nu}$ which is proportional to the Bach tensor of $g_{(0)}.$  
\end{remark}

From the equation $A_{\mu\nu}=B_{\mu\nu}$ at $y=y_{\text{br}}$ and the identity $(\ref{eq:K_rho_p}),$ equations for $\tilde{\rho}$ and $\tilde{p}$ can be obtained by considering
\begin{equation*}
\begin{split}
-\frac{A_{00}}{2 \gamma_{00}}+\frac{A_{11}}{2 \gamma_{11}}+\frac{A_{22}}{\gamma_{22}}
&=-\frac{B_{00}}{2 \gamma_{00}}+\frac{B_{11}}{2 \gamma_{11}}+\frac{B_{22}}{\gamma_{22}},\\
A_{00}&=B_{00},
\end{split}
\end{equation*}
which is equivalent to the Ricatti differential equations
\begin{equation}
\begin{split}
\partial_4 \tilde{\rho}&=\frac{\kappa_5^2}{3} \tilde{\rho}^2 -\frac{2}{l}\tilde{\rho}-\frac{4}{l} \rho,\\
\partial_4 \left(3 \tilde{p}+\tilde{\rho}\right)&=-\frac{\kappa_5^2}{6} \left(3 \tilde{p}+\tilde{\rho}\right)^2 -\frac{2}{l}\left(3 \tilde{p}+\tilde{\rho}\right)-\frac{4}{l}\left(3p+ \rho\right).\\
\end{split}
\label{eq:ricatti}
\end{equation}
Since it is widely believed that the theory of quantum gravity 
would have a minimal length scale, the effects of brane thickness should be included. 
To analyse these equations, it is assumed that a homogeneous brane is localised between $y=-y_0/2$ and $y=y_0/2$ (see e.g. \cite{Dzhunushaliev.Folomeev.ea_ThickBraneSolutions_2010}). Consequently, this density profile specifies $\partial_4 \rho=0$ inside the brane.
Therefore, the well-known transformation of the Ricatti differential equations to a second order linear equation yields
\begin{equation*}
\tilde{\rho}=-\frac{6}{l \kappa_4^2} \frac{c_1 z_1 \lambda_1+c_2 z_2 \lambda_2}{c_1 z_1 + c_2 z_2},
\end{equation*}  
where $\lambda_{1,2}=-1/l \pm \sqrt{1/l^2+2 \kappa_4^2 \rho/3},$ $z_{1,2}=\exp(\lambda_{1,2} y_{\text{br}})$ and $c_{1,2}$ are arbitrary constants. The solution
\begin{equation*}
3\tilde{p}+\tilde{\rho}=\frac{12}{l \kappa_4^2} \frac{d_1 z_1 \lambda_1+d_2 z_2 \lambda_2}{d_1 z_1 + d_2 z_2},
\end{equation*}  
of the second Ricatti equation follows in an analog manner and introduces the additional constants $d_1$ and $d_2.$

Consequently, equation $(\ref{eq:brane2})$ can be written without $\tilde{\rho}$ and $\tilde{p}.$ The analysis of this equation in the non-relativistic case is the aim of the next section. 

Before that, the setting $y_{\text{br}}=0$ is taken for the rest of this work because 
the agreement between the RBW model and the $\Lambda$CDM model was shown for branes at $y_{\text{br}}=0,$ (cf. \cite{Henke_Standard_Cosmology_2021}). In order to ensure a static spherically symmetric space-time in that case, the constraint
\begin{equation}
b^2(r,y)=\frac{r^2}{a(r,0)} a(r,y),
\label{eq:sym_sph_constr}
\end{equation}
is considered in the following.
Now, equation $(\ref{eq:brane2})$ leads to
\begin{equation*}
\frac{R_{00}}{2n}+\frac{R_{11}}{2a}+\frac{R_{22}}{r^2}=8 \pi G_4 \rho+4 \pi G_4 \tilde{\rho},
\end{equation*}
and therefore 
\begin{equation}
\left( \frac{r}{a} \right)'=1-\Lambda_5 r^2 -\kappa_4^2 \rho r^2 -\frac{\kappa_4^2}{2} \tilde{\rho} r^2.
\label{eq:sum_einstein}
\end{equation}
The integration of $(\ref{eq:sum_einstein})$ leads to
\begin{equation}
a^{-1}=1-\frac{\Lambda_5}{3}r^2 -\frac{2 G_4 M}{r}-\frac{G_4 \tilde{M}}{r},
\label{eq:inv_a}
\end{equation}
where $M(r)=4 \pi \int_0^r \rho(\xi) \xi^2\, d\xi$ and $\tilde{M}(r)=4 \pi \int_0^r \tilde{\rho}(\xi) \xi^2\, d\xi.$
Reinserting equation $(\ref{eq:inv_a})$ in $(\ref{eq:brane2})$ gives
\begin{equation}
\frac{n'}{2n}=\vartheta \frac{G_4 M}{r^2},
\label{eq:dn_n}
\end{equation}
where
\begin{equation}
\vartheta=\frac{1-\frac{\Lambda_5}{3 G_4 M}r^3+\frac{\tilde{M}}{2M}+\frac{4 \pi}{M}p r^3 +\frac{2 \pi}{M} \tilde{p} r^3}{1-\frac{\Lambda_5}{3}r^2 -\frac{2 G_4 M}{r}-\frac{G_4 \tilde{M}}{r}}. 
\label{eq:dn_n2}
\end{equation}
Using the setting $\Phi(r)=\ln(n(r))/2$, equation $(\ref{eq:dn_n})$ leads to the non-linear generalisation of the Newtonian Poisson equation
\begin{equation}
\nabla \cdot \left( \frac{1}{\vartheta} \nabla \Phi \right)
=\frac{1}{r^2} \frac{d}{dr} \left(\frac{r^2}{\vartheta} \frac{d}{dr} \Phi \right)
=4 \pi G_4 \rho.
\end{equation}

\section{Non-relativistic gravity on the brane}
In the non-relativistic case and for sufficient small values of $l$, the assumptions $p \ll \rho \ll \kappa_4^{-2} l^{-2}=\kappa_4^{-2} |\Lambda_5|/6$ and $G_4 M/r \ll 1$ are valid. 
\begin{remark}
The RBW model approximates the $\Lambda$CDM model with a small effective cosmological constant which can be inferred by cosmological observations. The 5 dimensional cosmological constant $\Lambda_5$ does not provide an additive contribution within the effective cosmological constant because the modified Friedmann equation of the RBW model contains the $\Lambda_5$ dependencies essentially in different denominator terms.
In \cite{Henke_Standard_Cosmology_2021} it has been demonstrated, that 
the greater $\Lambda_5,$ 
the better the approximation of the $\Lambda$CDM model will be.
Therefore, the assumption $\rho \ll \kappa_4^{-2} |\Lambda_5|/6$ can easily be fulfilled.
\end{remark}

Consequently, it follows 
\begin{equation}
\begin{split}
\tilde{\rho}& \approx \frac{12}{l^2 \kappa_4^2} \frac{1}{1+\frac{c_2}{c_1} },\\
3\tilde{p}+\tilde{\rho}&\approx-\frac{24}{l^2 \kappa_4^2} \frac{1}{1+\frac{d_2}{d_1}},\\
\tilde{M}&\approx \frac{2 r^3}{G_4 l^2 } \frac{1}{1+\frac{c_2}{c_1} }.
\end{split}
\label{eq:non_relativistic}
\end{equation} 
Then, equation $(\ref{eq:dn_n})$, the definition of $\Phi$ and the setting $e_1=1-1/(1+\frac{c_2}{c_1})$ leads to the line element
\begin{equation}
ds_4^2=-e^{2\Phi} dt^2+ \frac{1}{1-\frac{2 G_4 M}{r} -\frac{\Lambda_5}{3} r^2 e_1} dr^2 +r^2 d\Omega^2,
\label{eq:non_rel_metric}
\end{equation}
of the brane. Moreover, from the equations $(\ref{eq:dn_n}), (\ref{eq:dn_n2})$ and $(\ref{eq:non_relativistic})$ it follows that 
\begin{equation}
\Phi' =
\frac{1-\frac{\Lambda_5}{3 G_4 M}r^3 e_2}{1-\frac{\Lambda_5}{3}r^2 e_1}
\Phi_N',
\label{eq:mond1}
\end{equation}
where $e_2=1-1/(1+\frac{d_2}{d_1})$ and $\Phi_N=-G_4 M/r$ denotes Newton's potential.
Notice, that the characteristic MOND-scale $a_0$ doesn't appear in equation $(\ref{eq:mond1}).$ It is just another name for the second degree of freedom of $(\ref{eq:mond1}),$ which goes back to the solution of the Ricatti equation $(\ref{eq:ricatti}).$ 

Using the notations 
$a=-\Phi'$, $a_N=-\Phi_N'$, $a_0=1.2 \cdot 10^{-10} m s^{-2}$,
$k_1=2 G_4 M e_1/(l^2 a_0)$ 
and
$k_2=2\sqrt{G_4 M} e_2/(l^2 a_0^{3/2}),$ 
it follows that
\begin{equation}
a=\nu\left(\frac{a_N}{a_0}\right) a_N,\quad \nu(y)=\vartheta(r)=\frac{1+k_2 y^{-3/2}}{1+k_1 y^{-1}}.
\label{eq:mond2}
\end{equation}
The constants $e_1$ and $e_2$ encode boundary conditions from the galaxies to be considered and can be used for the individual fitting of the rotation curve data.
The quality of this BRAND (Brane Newtonian Dynamics) approach is independent of the concrete value of $l$ or $\Lambda_5,$ since it can be compensated with $e_1$ and $e_2.$ 
Equation $(\ref{eq:mond2})$ allows the investigation of two regimes:
\begin{equation}
\begin{split}
\nu(y) &\to 1 \text{ for } y \gg 1, \\
\nu(y) &\to \frac{k_2}{k_1}y^{-1/2} \text{ for } y \ll 1,
\end{split}
\label{eq:mond_constr}
\end{equation}
where the first regime specifies the usual Newtonian Dynamics and the second regime describes the modified Newtonian Dynamics.

If the mass can be approximated in the second regime where $G_4 M(r)/a_0 \ll r^2$ with the total mass of the galaxy $M(r) \approx M_{\infty}=\lim_{r \to \infty} M(r),$ one can ask for the agreement with the well known MOND hypothesis.
That is to say that the velocity of an object around a mass $M$ is in the Newtonain limit $v=\sqrt{G_4 M/r}$ and in the modified Newtonian regime $v=\sqrt[4]{a_0 G_4 M}.$
According to \cite{Famaey.McGaugh_Modified_Newtonian_Dynamics_2012}, this is the case when $k_2/k_1=1.$ Then, a short calculation yields the equivalent relationship 
\begin{equation*}
e_2=\sqrt{a_0 G_4M_{\infty}}e_1.
\end{equation*}
Notice that this relationship between $k_1$ and $k_2$ is not a necessary requirement of the method. 
The success of the MOND model suggests that this choice can often be observed for spherically symmetric galaxies with an isolated mass. 

Moreover, for $e_1>0$ the total mass approximation and the integration of equation $(\ref{eq:mond1})$ leads to the non-relativistic potential
\begin{equation*}
\begin{split}
\Phi&=-\frac{G_4 M_\infty}{r} + \frac{e_2 \log(1-\Lambda_5 e_1 r^2/3)}{2 e_1}\\ 
&- \sqrt{-\Lambda_5 e_1/3} G_4 M_\infty \arctan(\sqrt{-\Lambda_5 e_1/3}r)
\end{split}
\end{equation*}
of the RBW model. The case $e_1<0$ gives a $\log$-term with a negative argument and is therefore not physical.

In more general cases, when there is no isolated mass with a natural asymptotic mass definition $M_\infty$, the explicit mass distribution $M$ can be used and the constants $e_1$ and $e_2$ can be determined by rotation curves of galaxies. In these situations, the function $\nu$ from $(\ref{eq:mond2})$ does not provide a universal decription such as MOND. Future research must show whether nature allows a universal description at this point.

\begin{remark}
This section ensures that the dynamics of the RBW model reproduce the correct amount of \glqq missing mass\grqq. It would be very interesting if the correct magnitude of  \glqq missing mass\grqq \, is also considered from the gravitational lensing point of view. 
In order to ensure that the RBW model explains the observations of intergalactic lensing without dark matter, 
it is required that $\psi \approx -\varphi$ (cf. \cite{Famaey.McGaugh_Modified_Newtonian_Dynamics_2012}), where $\varphi$ and $\psi$ are the potentials of the isotropic metric
\begin{equation}
ds_4^2=-e^{2\varphi} dt^2+ e^{2\psi}\left( d\rho^2 +\rho^2 d\Omega^2\right).
\label{eq:non_rel_isometric}
\end{equation}
To adopt this metric, the transformation from  
$(\ref{eq:non_rel_metric})$ to $(\ref{eq:non_rel_isometric})$ leads to the identities 
\begin{equation} 
\frac{1}{1-s(r)} dr^2 = e^{2 \psi(\rho)} d\rho^2, \quad r=e^{\psi(\rho)}\rho,
\label{eq:trans_id}
\end{equation}
where the setting $s(r)=2 G_4 M_\infty/r +\Lambda_5 r^2 e_1/3$ is used. 
Dividing both equations and satisfying $\rho \to \infty$ for $r \to \infty,$ it follows that
\begin{equation*}
\frac{1}{\sqrt{1-s(r)}} \frac{dr}{r}=\frac{d\rho}{\rho}.
\label{eq:trans_integrand}
\end{equation*}
Unfortunately, the integration of the last equation is difficult. Even an approximate analysis with $|s(r) |\ll 1$ is not possible in the \glqq deep MOND\grqq\, regime, since 
\begin{equation*}
-\frac{\Lambda_5}{3} r^2 e_1=k_1 y^{-1} \gg 1 \text{ for } y \ll 1.  
\end{equation*}
To understand gravitational lensing phenomena further research is needed for the RBW model. 
\end{remark}

\section{Concluding remarks}
In this paper, it has been demonstrated that the RBW model reproduce the correct amount of \glqq missing mass\grqq\, and satisfies Milgrom's law in special cases. Together with the previous paper  
\cite{Henke_Standard_Cosmology_2021}, it is shown that the RBW model is the first approach that includes 
the MOND hypothesis 
and the $\Lambda$CDM model without the cosmological constant problem. 

To ensure that the RBW model is as successful as Milgrom's theory, further research should extend the RBW model to asymmetric situations. 

The “missing matter” explanation of the RBW model requires an inhomogeneous metric. Therefore, it is not surprising that the RBW model with the homogeneous metric of Robertson and Walker contains no explanation of the cosmological “missing matter“ (see \cite{Henke_Standard_Cosmology_2021}). However, there is a reasonable hope that further investigations of the RBW model with a suitable inhomogeneous metric can explain the cosmological ”missing matter“.

\appendix
\section{Equations on the brane}
In this section we review the basic aspects of timelike hypersurfaces.
Following the notation of \cite{Aliev.Guemruekcueoglu_Gravitational_field_2010}, we introduce a $(4+1)$ decomposition of a five dimensional bulk space-time in the spirit of Arnowitt, Deser and Misner (ADM) \cite{Arnowitt.Deser.ea_Dynamics_general_relativity_1962}. 
Let $X^A$ be the coordinates of the metric $G_{AB},\,A,B=0,1,\dots,4$ and 
$y=y(X^A)$ a scalar function such that $y=\text{const}$ describes a family of non-intersecting $(3+1)$-dimensional hypersurfaces $\Sigma(y).$ The boundary which is denoted by $\Sigma_D=\Sigma(y_{\text{br}}).$

The normal vector field $n_A$ of the hypersurface with $n_A n^A=1$ can be introduced by $n_A=N \partial_A y,$ where $N=|G^{AB} \partial_A y \partial_b y|^{-1/2}$ is the lapse function.

The coordinates $X^A=X^A(x^\mu,y), \mu=0,\cdots,3$ can be parameterised in terms of the intrinsic hypersurface $x^\mu=(t,x^i),\, i=1,\cdots,3$ and the fifth coordinate $y.$ The change of $X^A$ with respect to this parametrisation is given by 
\begin{equation*}
dX^A=y^A dy +e_\mu^A dx^\mu,
\end{equation*}  
where 
\begin{equation*}
y^A=N n^A +N^\mu e_{\mu}^A 
\end{equation*}
is the evolution vector into the fifth dimension and $N^\mu$ is the shift vector. 
\begin{figure}[!htbp]
\begin{center}
\includegraphics[width=1.0\columnwidth]{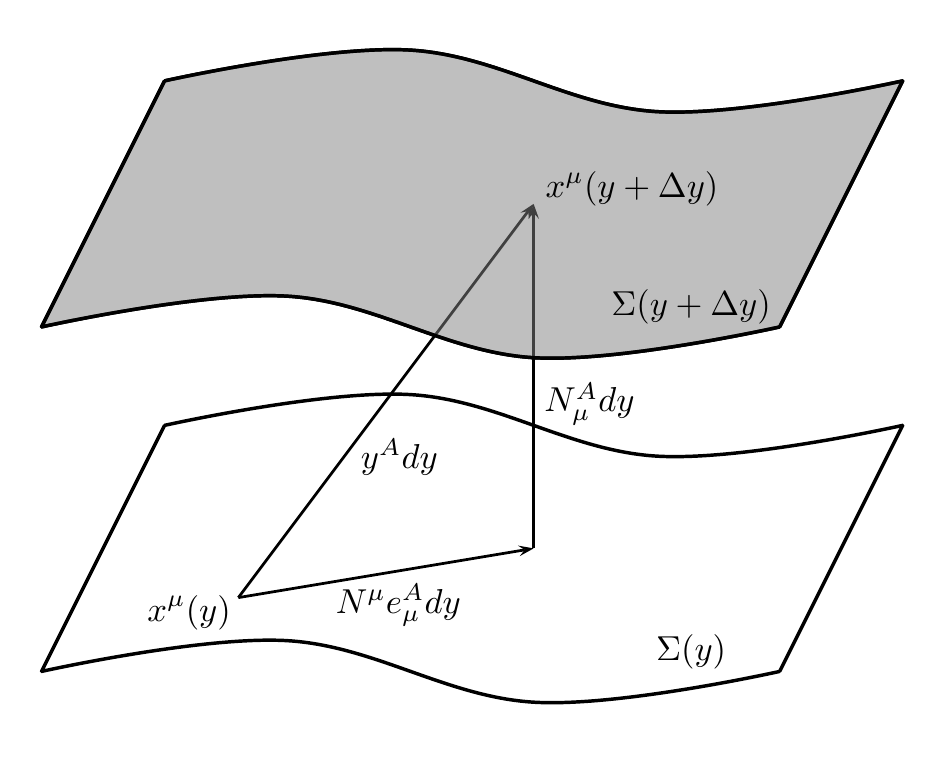}
\caption{Foliation of the $4+1$ dimensional space-time into hypersurfaces at constant $y.$}
\label{fig:branes}
\end{center}
\end{figure}
Then, the local frame is given by 
\begin{equation*}
e^A_\mu=\frac{\partial X^A}{\partial x^\mu}=\delta_\mu^A,\quad
y^A=\frac{\partial X^A}{\partial y}=\delta_y^A.
\label{}
\end{equation*} 
Now, the metric on the hypersurface has the form
\begin{equation*}
\gamma_{AB}=G_{AB}-n_A n_B, \quad \gamma_{\mu \nu}=\gamma_{AB} e^A_\mu e^B_\nu.
\label{}
\end{equation*}
The ADM-like decomposition can be introduced by 
\begin{equation}
G_{AB}=
\begin{pmatrix}
\gamma_{\mu \nu}& N_\mu     \\
 N_\nu & N^2 + N_\mu N^\mu   \\
\end{pmatrix}.
\label{eq:G_ADM}
\end{equation}
By computing the metric determinant and the inverse metric, it follows that $\sqrt{-G}=N\sqrt{-\gamma}$ and
\begin{equation*}
G^{AB}=
\begin{pmatrix}
\gamma^{\mu\nu} + N^\mu N^\nu/N^2 & -N^\mu/N^2 &  \\
 - N^\nu/N^2& 1/N^2\\
\end{pmatrix}.
\label{}
\end{equation*}
With these notations at hand, the unit normal vector $n_A$ is given by
\begin{equation*}
n_A= \left(0,0,0,0,N\right)^T, \, n^A=\left(-\frac{N^\mu}{N},\frac{1}{N}\right)^T.
\label{}
\end{equation*}
Moreover, the extrinsic curvature is defined by
\begin{equation}
K_{\mu\nu}=\frac{1}{2} (\mathcal{L}_n G_{AB}) e_\mu^A e_\nu^B
= \frac{1}{2N}(\partial_4 \gamma_{\mu \nu} -\nabla_\mu N_\nu-\nabla_\nu N_\mu),
\label{eq:Kmunu}
\end{equation}
where $\mathcal{L}_n$ is the Lie derivative of the unit normal vector $n_A$
and $\nabla_\mu$ denotes the covariant derivative operator w.r.t. the metric $\gamma_{\mu\nu}.$
Notice that the surface energy-momentum tensor $\tilde{T}_{\mu\nu}$ provides also a representation for the extrinsic curvature
\begin{equation}
K_{\mu\nu}=-\frac{\kappa_5^2}{2}\left( \tilde{T}_{\mu\nu}- \frac{1}{3}\gamma_{\mu\nu} \tilde{T} \right).
\label{eq:K_T_matt}
\end{equation}
Now, the usual definition of the energy-momentum tensor 
\begin{equation*}
\tilde{T}_{\mu\nu}=(\tilde{\rho} + \tilde{p}) u_\mu u_\nu + \tilde{p} \gamma_{\mu\nu},\quad u^\mu=\left( (-\gamma_{00})^{-1/2},0,0,0 \right),
\label{}
\end{equation*}
yields 
\begin{equation}
K_{\mu\nu}=-\frac{\kappa_5^2}{2}\left( \frac{\tilde{\rho}}{3} \gamma_{\mu\nu}+ \left(\tilde{\rho}+\tilde{p}\right) u_\mu u_\nu \right).
\label{eq:K_rho_p}
\end{equation}

According to \cite{Aliev.Guemruekcueoglu_Gravitational_field_2010}, the five dimensional Ricci tensor and Ricci scalar can be split into intrinsic and extrinsic surface terms
\begin{equation}
\begin{split}
\li{R_{\mu\nu}}&=R_{\mu\nu}-\frac{1}{N} \left\{ \left( \partial_4 - \mathcal{L}_N \right) K_{\mu\nu} + \nabla_{\mu}\nabla_{\nu}\right\} \\
&+ 2 K_{\mu}^\alpha K_{\alpha \nu}-K K_{\mu\nu},
\quad K=\gamma^{\mu\nu} K_{\mu\nu},
\end{split}
\label{eq:Ricci_ten_5}
\end{equation}
and
\begin{equation}
\li{R}=R-\frac{2}{N}\left( \partial_4 K -\mathcal{L}_N + \Box N \right) 
- K_{\mu \nu}K^{\mu\nu}-K^2,
\label{eq:Ricci_sca_5}
\end{equation}
where $\Box=\nabla_{\mu} \nabla^{\mu}$ and $\mathcal{L}_N$ is the Lie derivative w.r.t. the shift vector $N^\mu.$

Solving the five dimensional Einstein equation in a vacuum region 
\begin{equation*}
\li{\pi}_{AB}=0,\; \text{in } \mathcal{M},
\end{equation*}
where
\begin{equation*}
\li{\pi}_{AB}= \li{R_{AB}}-\frac{1}{2}G_{AB} \li{R}+\Lambda_5 G_{AB}, 
\end{equation*}
leads to an asymptotic AdS space-time with $\Lambda_5=-6/l^2$ and $\li{R}=-20/l^2.$
In order to obtain an asymptotic expansion of $G_{AB}$ near the conformal boundary, the coordinates of Fefferman-Graham  
\begin{equation}
ds^2=\frac{l^2}{r^2}\left( g_{\mu\nu}(x,r)\, dx^\mu dx^\nu +dr^2\right),\quad r=l e^{y/l}.
\label{eq:coord_Feff_Gra}
\end{equation}
with the associated expansion
\begin{equation}
\begin{split}
g(x,r)&=g_{(0)}(x)+r^2 g_{(2)}+r^4 g_{(4)}(x)\\
&+r^4 \log(r^2)h_{(4)}(x)+ \mathcal{O}(r^{6})
\end{split}
\label{eq:FG}
\end{equation}
are used. Notice that the metric of the boundary $g_{(0)}$ is only defined up to a conformal transformation.

With the help of this asymptotic expansion (cf. \cite{de_Haro.Skenderis.ea_Holographic_Reconstruction_2001})
one can isolate a finite number of terms that diverge on the conformal boundary.  The subtraction of these counterterms is called renormalisation and can be written as
\begin{equation*}
S=S_{\text{gr}}+S_{\text{ct}}
\end{equation*}
where
\begin{equation}
\begin{split}
S_{\text{gr}}&=\frac{1}{2\kappa_5^2} \int_{\mathcal{M}} \left( \li{R}-2\Lambda_5 \right) \sqrt{-G}\,d^5 x\\
&-\frac{1}{\kappa_5^2} \int_{\partial \mathcal{M}} K\sqrt{-\gamma}\, d^4 x, 
\end{split}
\label{eq:einstein_hilbert}
\end{equation}
is the usual Einstein-Hilbert action including the Gibbons-Hawking boundary term and 
\begin{equation*}
\begin{split}
S_{\text{ct}}&=-\frac{l}{2 \kappa_5^2} \int_{\partial \mathcal{M}}\left(\frac{6}{l^2} +\frac{R}{2}
 \right) \sqrt{-\gamma}\, d^4x\\
	&-\frac{1}{2 l\kappa_5^2} \int_{\partial \mathcal{M}} a_{(4)} r_{\text{br}}^4\log\left(r_{\text{br}}^2 \right)\sqrt{-\gamma}\, d^4x,\\
\end{split}
\label{eq:Sct}
\end{equation*}
is the counterterm which cancels the divergent terms in $S_{\text{gr}}.$ 
The logarithmic term originates from the integration of the bulk integral $\int_{\mathcal{M}} \cdots d^5$ in $S_\text{gr}$
and leads to the coefficient
\begin{equation*}
a_{(4)}=\frac{1}{2} \text{tr}\left( g_{(0)}^{-1}g_{(2)} \right)^2 -\frac{1}{2} \text{tr}\left( \left[ g_{(0)}^{-1} g_{(2)} \right]^2 \right).
\end{equation*}
Notice, that the last term of $S_{\text{ct}}$ makes explicit reference to the cut-off $r_{\text{br}}=le^{y_{\text{br}}/l}.$ Due to the expansion $(\ref{eq:FG}),$ the cut-off should be close to the conformal boundary at $r=0.$


The finiteness of $\delta S$ defines a valid variational principle 
\begin{equation}
\delta S = -\frac{1}{2} \int_{\partial M} T_{\mu\nu} \delta \gamma^{\mu\nu} \sqrt{-\gamma}\, d^4 x,
\label{eq:dS}
\end{equation}
where
\begin{equation*}
T_{\mu\nu}= - \frac{2}{\sqrt{-\gamma}} \frac{\delta S_m}{\delta \gamma^{\mu\nu}},\\
\end{equation*}
and
\begin{equation*}
S_m=\int_{\partial \mathcal{M}} L_{\text{matter}} \sqrt{-\gamma}\, d^4 x 
\end{equation*}
are the matter terms with the Lagrangian density $L_{\text{matter}}.$

Now, standard calculations of equation $(\ref{eq:dS})$ gives the equivalent formulation
\begin{equation}
\begin{split}
& \frac{1}{2\kappa_5^2} \int_{\mathcal{M}} \li{\pi}_{AB}  \delta G^{AB} \sqrt{-G}\, d^5 x\\
&+\frac{1}{2\kappa_5^2} \int_{\partial \mathcal{M}}\pi_{\mu\nu} \delta \gamma^{\mu\nu} \sqrt{-\gamma}\, d^4 x=0.
\end{split}
\label{eq:variation}
\end{equation}
Here, the following notations have been adopted
\begin{equation}
\begin{split}
\pi_{\mu\nu}&= \frac{l}{2}\bigg( E_{\mu\nu}-\frac{6}{l^2} \gamma_{\mu\nu} -4 r_{\text{br}}^2\log\left(r_{\text{br}}^2\right) h_{(4)\mu\nu}\bigg)\\
&-\kappa_5^2 \bigg( \frac{\tilde{T}_{\mu\nu}[\gamma]}{2}+ T_{\mu\nu}[\gamma]\bigg),\\
E_{\mu\nu} &=R_{\mu\nu}[\gamma] -\frac{1}{2}R[\gamma] \gamma_{\mu\nu}.
\end{split}
\label{eq:not_einstein}
\end{equation}
If one applies dynamical metrices $\delta \gamma_{\mu\nu}\neq 0,$ equation $(\ref{eq:variation})$ is satisfied by
\begin{equation}
\begin{split}
\li{\pi}_{AB}&=0,\; \text{in } \mathcal{M},\\
\pi_{\mu\nu}&=0,\; \text{in } \partial\mathcal{M}. 
\end{split}
\label{eq:einstein}
\end{equation}

Using the equations $(\ref{eq:Ricci_ten_5})$ and $(\ref{eq:Ricci_sca_5}),$ the first equation of $(\ref{eq:einstein})$ can be written as 
\begin{equation}
\begin{split}
&E_{\mu\nu}+\Lambda_5 \gamma_{\mu\nu}+\frac{1}{N}\left(\gamma_{\mu\nu} \Box - \nabla_\mu \nabla_\nu \right)N \\
=&\frac{1}{N}\left\{ \left( \partial_4 -\mathcal{L}_N \right) \left( K_{\mu\nu}-\gamma_{\mu\nu}K \right) \right\} \\
+& 3K K_{\mu\nu}
-2 K_{\mu}^\alpha K_{\alpha\nu}
-\frac{1}{2} \gamma_{\mu\nu} \left( K^2+ K_{\alpha\beta}K^{\alpha\beta} \right).
\end{split}
\label{eq:einstein4}
\end{equation}
Moreover, to fulfil the five dimensional Einstein equation, the following constraints has to be satisfied
\begin{equation}
\begin{split}
-\frac{1}{2} \left( R-K^2 +K_{\alpha\beta}K^{\alpha\beta} \right)+\Lambda_5
&=0,\\ 
 \nabla_\nu K_\mu^\nu -\nabla_\mu K
&=0,
\end{split}
\label{eq:constraints}
\end{equation}
whereas the second equation implies the covariant conservation of $\tilde{T}_{\mu\nu}$
\begin{equation}
\nabla^\nu \tilde{T}_{\mu\nu}=0.
\label{eq:matter_cons}
\end{equation}
Now, it is supposed that the boundary metric $g_{(0)}$ has a vanishing Bach tensor. Then, using the property that the Bach tensor is proportional to $h_{(4)\mu\nu}$ (see \cite{Genolini.Cassani.ea_Holographic_renormalization_2017}), the following Einstein equation 
\begin{equation}
E_{\mu\nu}+\Lambda_5 \gamma_{\mu\nu}=
\kappa_4^2 \left(T_{\mu\nu}+\frac{1}{2} \tilde{T}_{\mu\nu}\right), \quad \kappa_4^2=\frac{2\kappa_5^2}{l},
\label{eq:en_ten_cft}
\end{equation}
is equivalent to the second equation of $(\ref{eq:einstein}).$

Suppose from now on that the shift vector $N_\mu$ vanishes. Applying the definition $(\ref{eq:Kmunu})$ and the identity $(\ref{eq:K_rho_p}),$ the following constraints follows for $i,j=1,2,3.$
\begin{equation*}
\partial_4 \ln \gamma_{00}=N \kappa_5^2 \left( \frac{2}{3} \tilde{\rho} +\tilde{p}\right),\quad
\partial_4 \ln \gamma_{ij}=-N \kappa_5^2 \frac{\tilde{\rho}}{3}, 
\end{equation*}
where the last equation implies the metric constraints (cf. also the constraint (\ref{eq:sym_sph_constr})) 
\begin{equation}
\begin{split}
\gamma_{22}(x^\mu,y)&=C_2(x^\mu) \gamma_{11}(x^\mu,y),\\
\gamma_{33}(x^\mu,y)&=C_3(x^\mu) \gamma_{11}(x^\mu,y).
\end{split}
\label{eq:metric_constr}
\end{equation}



\end{document}